\documentclass[usenatbib]{mn2e}
\usepackage{graphicx}
\usepackage{subfigure}
\usepackage[colorlinks=true,linkcolor=black,citecolor=black,urlcolor=blue]{hyperref}
\usepackage{enumerate}
\usepackage{amsmath}
\usepackage{mathtools}
\usepackage{IEEEtrantools}
\usepackage{float}
\usepackage{txfonts}
\usepackage[applemac]{inputenc}
\usepackage[english]{babel}
\usepackage{booktabs}
\usepackage{multirow}
\usepackage{amstext}
\usepackage{subfigure}
\usepackage{graphicx}
\usepackage[np,noautolanguage]{numprint}
\usepackage{url}
\usepackage{hyperref}
\usepackage{pdfsync}
\usepackage{threeparttable}
\usepackage{color}
\usepackage[T1]{fontenc}{
\usepackage{aecompl}

\newcommand{\apjl}{ApJ}
\newcommand{\apjs}{ApJS}
\newcommand{\ao}{Appl.~Opt.}
\newcommand{\aap}{A\&A}

\newcommand{\beq}{\begin{equation}}
\newcommand{\eeq}{\end{equation}}

\newcommand{\msun}{\,{\rm M_\odot}}
\definecolor{grey}{rgb}{0.5,0.6,0.7}
\def \simlt { \lower .75ex \hbox{$\sim$} \llap{\raise .27ex \hbox{$<$}} }
\definecolor{purple}{rgb}{0.65,0.15,0.9}
\definecolor{darkorange}{rgb}{0.8,0.3,0}
\definecolor{olive}{rgb}{0.4,0.6,0.25}
\definecolor{darkgreen}{rgb}{0,0.7,0}
\definecolor{darkred}{rgb}{0.5,0,0}

\title[``Direct collapse"  black hole seeds]{On the number density of  ``direct collapse"  black hole seeds}
\author[Habouzit et al.]{M\'{e}lanie Habouzit$^{1}$\thanks{E-mail: habouzit@iap.fr},
 Marta Volonteri$^{1}$,
  Muhammad Latif$^{1}$,
  Yohan Dubois$^{1}$,
  \newauthor
 and S\'{e}bastien Peirani$^{1}$\\
$^1$Institut d'Astrophysique de Paris, Sorbonne Universit\'{e}s, UPMC Univ Paris 6 et CNRS, UMR 7095, 98 bis bd Arago, 75014 Paris, France\\}

\begin{document}
\maketitle

\begin{abstract}
Supermassive black holes (BHs) reside in the center of most local galaxies, but they also power active galactic nuclei and quasars, detected up to $z=7$. These quasars put constraints on early BH growth and the mass of BH seeds.
The scenario of ``direct collapse'' is appealing as it leads to the formation of large mass BH seeds, $10^4-10^6 \rm{M_{\odot}}$, which eases explaining how quasars at $z=6-7$ are powered by BHs with masses $>10^9 \msun$. 
Direct collapse, however, appears to be rare, as the conditions required by the scenario are that gas is metal-free, the presence of a strong photo-dissociating Lyman-Werner flux, and large inflows of gas at the center of the halo, sustained for $10-100$ Myr.  
We performed several cosmological hydrodynamical simulations that cover a large range of box sizes and resolutions, thus allowing us to understand the impact of several physical processes on the distribution of direct collapse BHs. We identify halos where direct collapse can happen, and derive the number density of BHs.  We also investigate the discrepancies between hydrodynamical simulations, direct or post-processed, and semi-analytical studies. Under optimistic assumptions, we find that for direct collapse to account for BHs in normal galaxies, the critical Lyman-Werner flux required for direct collapse must be about two orders of magnitude lower than predicted by 3D simulations that include detailed chemical models.  However, when supernova feedback is relatively weak, enough direct collapse BHs to explain $z=6-7$ quasars can be obtained for Lyman-Werner fluxes about one order of magnitude lower than found in 3D simulations.

\end{abstract}


\section{Introduction}
\label{sec:intro}
The formation of supermassive black holes (BHs) is still an unresolved problem. It is now well established that most local galaxies harbor a supermassive BH in their center, with mass even above $10^{10} \rm{M_{\odot}}$ for the most massive ones \citep{2011Natur.480..215M}.  
We observe quasars, the active tail of the BHs distribution, up to redshift $z>6$ \citep{fan06,jiange09,mortlock11}. Simulations suggest that these high-redshift quasars essentially grow by direct cold filamentary infall~\citep{dimatteoetal12,duboisetal12} and their large masses require that BH seeds must have been formed at very early times, well before $z=6$, in order to acquire $10^{9}\, \rm{M_{\odot}}$ within a span of less than 1 Gyr.

A popular channel to form BH seeds  has been as remnant of the first generation of stars, namely PopIII stars \citep{Madau2001,Volonteri2003}. BHs are predicted to form in metal-free mini-halos ($M_{\rm{h}} \approx 10^{5} \rm{M_{\odot}}$) at redshifts $z=20-30$ from the remnants of PopIII stars, which are stars of primordial composition.  Studies suggest that the mass of these PopIII stars could range from 10 to 1000 $\rm{M_{\odot}}$ \citep[e.g.,][]{Bromm_Yoshida_2011,Hirano14}.  A sufficiently massive PopIII star ($> 260 \, \rm{M_{\odot}}$) can collapse and form a BH retaining $\sim$ half the mass of the star, leading to a population of $\sim 100\, \rm{M_{\odot}}$ BHs \citep{Fryer01}. This has been a major criticism of the scenario, because starting from this low-mass seed population, it means that such seed BHs must grow at a rate close to the maximum Eddington rate their entire life to explain the masses of $z>6$ quasars. 

One recent way of solving the problem of the formation of such massive objects is by the  direct collapse of pristine gas triggered by dynamical processes  \citep{Loeb1994,Eisenstein1995,Koushiappas2003,Begelman2006,Lodato2006,Mayer2010}, or invoking isothermal collapse in primordial halos \citep{Bromm2003,spaans2006,Dijkstra2008,Latif2013} . This scenario has become very popular, as it can lead to the formation of $10^{4}-10^{6}\, \rm{M_{\odot}}$ seeds, making it easier  to reproduce the quasar population at  $z>6$.

The physical conditions (listed below) required for isothermal collapse in primordial halos, which has become the most popular version of direct collapse  in the last couple of years, and whose BH seeds are typically referred to as `DCBHs', are numerous, and make the channel a rare event. First, one needs a halo that has reached the atomic cooling threshold, $\sim 10^7-10^8 \msun$, but it is still pristine, i.e., metal-free. Second, one needs that the molecular hydrogen formation has been suppressed throughout the halo's evolution. When these conditions are fulfilled, the gas in the halo can collapse isothermally and without fragmenting in the absence of efficient coolants, namely metals and molecular hydrogen. Conversely, the presence of metals and molecular hydrogen would decrease the temperature of the gas, and so the Jeans mass. This could lead to the fragmentation of the gas cloud, therefore the formation of only one massive object is unlikely, and the formation of several less massive objects, namely PopIII stars, is instead expected. The destruction and prevention of  molecular hydrogen can be accomplished by strong photo-dissociating radiation (Lyman-Werner, LW, photons with energy between 11.2 eV and  13.6 eV). A large inflow rate of gas at the center of the halo, higher than $0.1\, \rm{M_{\odot}/yr}$, and sustained for at least 10~Myr, is also needed to form a supermassive star-like object in the nucleus \citep{Begelman2006,Begelman2007b,Begelman2010,Ball2011,Hosokawa12,Hosokawa2013,2013A&A...558A..59S}. The BH mass can be up to 90\% of the stellar mass. 

Several aspects, on different physical scales, of the direct collapse scenario have been addressed in the last years.
However, so far, studies have focused either on small scale simulations to capture the physical processes leading to the gas collapse, or on semi-analytical studies to derive a  DCBH number density.
For instance, \citet{Latif2013} used zoomed cosmological hydrodynamical simulations of single halos to show that when all the conditions listed above are met, collapse can happen \citep[see also][]{Regan2014}. \citet[][A12 thereafter]{Agarwal2012SAM} and \citet[][A14 thereafter]{Agarwal2014},  investigated in post-processing spatial variations in the LW radiation and the importance of the clustering of the LW photon sources, using a 4 comoving Mpc (cMpc) cosmological hydrodynamical simulation. In the meanwhile, semi-analytical studies also derived the number densities of DCBHs, for example \citet[][D08 thereafter]{Dijkstra2008} computed the probability distribution function of the LW radiation that irradiates halos at redshift $z=10$, and showed that a small fraction of halos, $10^{-8}$ to $10^{-6}$, may be exposed to radiation higher than 20 times the background radiation \citep[see also][D14 thereafter]{Dijkstra2014}.   \citet{2015MNRAS.450.4350I} include the impact of X-rays, and predict a decrease in the formation rate of DCBHs per unit volume with redshift.
\cite{2015arXiv151001733A}  and \cite{2015arXiv151201111H} use semi-analytical models with merger trees based on the Extended Press \& Schechter formalism.  \citet{2015arXiv150705971H} develop a hybrid model, where they ``paint" galaxies, using the prescriptions of D08 and D14, over a dark matter only simulation, in order to include self-consistently the clustering of halos.  

In this paper, we use a set of three different cosmological hydrodynamical simulations, with increasing box sizes, allowing us to capture the physical processes on small scales in a small-volume, high-resolution simulation, to derive the number density of BHs and to test the impact of the SN feedback using a larger simulation  with intermediate resolution. Finally, a large-volume simulation, with lower resolution, Horizon-noAGN (Peirani et al, in prep), is also used to test whether the direct collapse scenario is able to explain the population of quasars that we observe at redshift $z=6$. We follow the approach introduced by D08, A12, D14 and A14 and model the LW radiation field on top of all the different simulations. A DCBH region finder code is applied to compute the DCBH number density function on all the simulations, for different redshifts. 

The paper is organized as follows. 
We first describe, in Section 2, the simulations we use in this work, and the modeling of the LW radiation intensity, in Section 3. In Section 4,  we investigate SN feedback and, in Section 5 how it can alter the number density of DCBH regions derived from simulations.
Section 6 is dedicated to the simulation Horizon-noAGN, for which we derive the number density of DCBH regions, and we investigate the feasibility of the DC scenario in explaining the population of quasars at $z=6$.
Section 7 explores a comparison between the approaches to derive the DCBH number density (which leads to discrepancies of several orders of magnitude), using either semi-analytical methods, or using hydrodynamical simulations.
Section 8 summarizes the results of this work.

\section{Simulation set up}
\begin{figure}
\centering
   \includegraphics[scale=0.45]{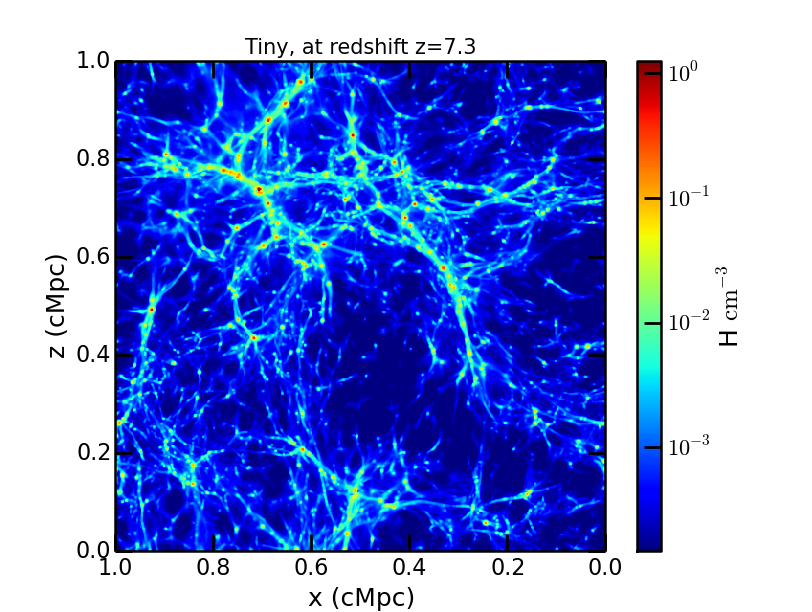}
   \includegraphics[scale=0.45]{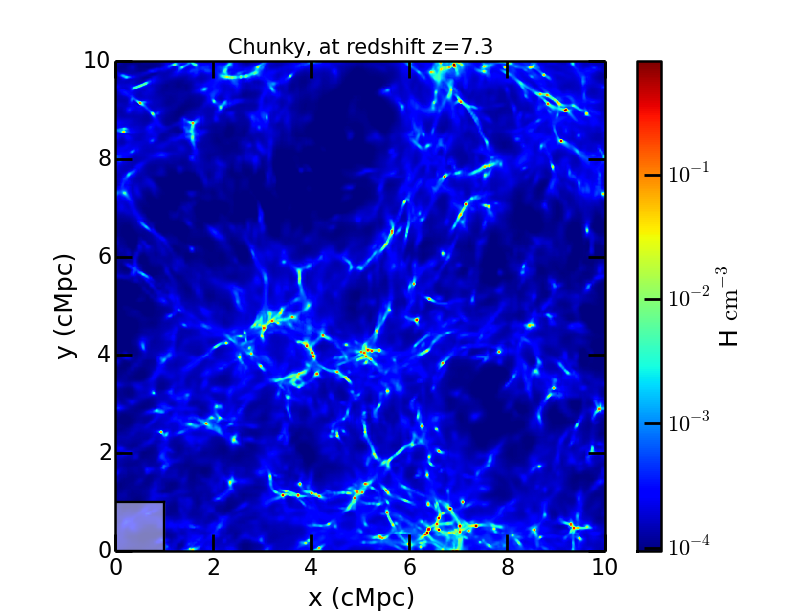}
   \includegraphics[scale=0.45]{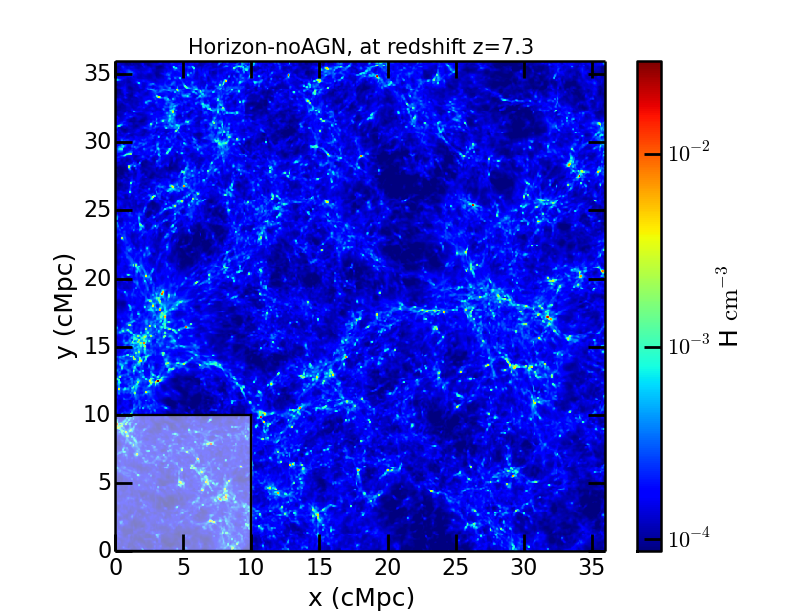}
\caption{Gas density maps representing the three simulations used in this work: Tiny ($\rm{L}_{\rm{box}}= 1$ cMpc, top panel), Chunky ($\rm{L}_{\rm{box}}=10$ cMpc, middle panel), and Horizon-noAGN ($\rm{L}_{\rm{box}}= 142$ cMpc, here we only show a small part of the simulation, a box of $\sim$ 40 cMpc side, bottom panel). The white squares mark the size of the previous simulation: in the gas density map of Chunky (middle panel), we show the size of Tiny, and in Horizon-noAGN map (bottom panel), we show the size of Chunky.}
\label{fig:density_maps}
\end{figure}

We have performed a set of simulations with increasing box sizes from 1 cMpc to 142 cMpc, using the adaptive mesh refinement hydrodynamical cosmological code {\sc ramses} \citep{Teyssier02}. 
Particles are projected on the grid with a cloud-in-cell interpolation and the Poisson equation is solved with an adaptive Particle-Mesh solver.
The Euler equations are solved with a MUSCL scheme using an approximate Harten-Lax-Van Leer Riemann solver, with a Min-Mod total variation scheme to interpolate the cell-centered values to their edge locations.
Cells are refined (unrefined) based on a quasi-Lagrangian criterion: with more (less) than 8 DM particles in a cell, or with a total baryonic mass higher (smaller) than 8 times the DM mass resolution.  
We summarize in the following the main characteristics of these simulations. Table~\ref{comp_simu} establishes a comparison between all the simulations parameters used in this work.
Fig.~\ref{fig:density_maps} shows gas density maps of our simulations Tiny, Chunky, and Horizon-noAGN. In the density map of Chunky (middle panel), the white square indicates the size of Tiny, and in the Horizon-noAGN map (bottom panel, only a small part of the simulation box is shown here), we show the size of the simulation Chunky.

\subsection{Simulation Tiny}
The smallest simulation, Tiny, is performed in a periodic box of $\rm{L}_{\rm{box}}= 1\, \rm{cMpc}$ size length with $256^{3}$ dark matter particles, corresponding to a mass resolution of $M_{\rm{DM,res}}=2082\, \rm{M_{\odot}}$. The simulation uses a $\Lambda$ cold dark matter cosmology, with total matter density $\Omega_{m}=0.276$, dark matter energy density $\Omega_{\Lambda}=0.724$, amplitude of the matter power spectrum $\sigma_{8}=0.811$, spectral index $\rm{n_{s}}=0.961$, baryon density $\Omega_{b}=0.045$ and Hubble constant $\rm{H_{0}}= 70.3 \, \rm{km\,  s^{-1} \, Mpc^{-1}}$. 
Simulations are run with 10 levels of refinement ($\ell_{\rm min}=8$ defines the number of cells on the coarse level, $\ell_{\rm max}=17$ defines the finest level of refinement), leading to a spatial resolution of 7.6 pc.  A new refinement level is allowed only when the expansion factor doubles, namely for $\rm{a_{exp}}=0.1,0.2,0.4$ and so on.

\subsection{Simulation Chunky}
Two intermediate volume simulations, Chunky, are also run, they differ by the model of SN feedback, in one case we use a ``thermal" SN feedback and in the other a ``delayed cooling'' SN feedback (both SN feedback models are described in Section 2.5.).  They use the same cosmology as Tiny. Simulations are performed in a periodic box of side $\rm{L}_{\rm{box}}= 10\, \rm{cMpc}$ with $128^{3}$ dark matter particles, corresponding to a mass resolution of $M_{\rm{DM,res}}=1.6 \times10^{7} \rm{M_{\odot}}$. These simulations are run on 11 levels of refinement ($\ell_{\rm min}=7$, $\ell_{\rm max}=17$), leading to a spatial resolution of 76.3 pc.

\subsection{Simulation Horizon-noAGN}
We use the Horizon-noAGN simulation (Peirani et al, in prep.), which is a version without BHs and AGN feedback of the Horizon-AGN simulation \citep{2014MNRAS.444.1453D}. This simulation has $\Omega_{\Lambda}=0.728$, $\Omega_{\rm{m}}=0.272$, $\Omega_{\rm{b}}=0.045$, $\sigma_{8}=0.81$, $\rm{n_{s}}=0.967$, and $\rm{H}_{0}=70.4\, \rm{km \, s^{-1} \, Mpc^{-1}}$. Simulations were run with $1024^{3}$ dark matter particles in a $\rm{L}_{\rm{box}}= 142 \,\rm{cMpc}$ size box, leading to a dark matter mass resolution of $M_{\rm{DM,res}} = 8 \times 10^{7} \rm{M}_{\rm{\odot}}$.  Simulations are run on 8 levels of refinement ($\ell_{\rm min}=10$, $\ell_{\rm max}=17$), leading to a spatial resolution of $\sim 1$ kpc.

\begin{table}
\caption{Simulation parameters for the four simulations used in this paper: Tiny, the two simulations Chunky, and Horizon-noAGN.}
\begin{center}
\begin{tabular}{llll}
\hline
\hline
Simulations & Tiny & Chunky & Horizon-noAGN \\
\hline
$\rm{L}_{\rm{box}}\, \rm{(Mpc)}$& 1 & 10 & 142 \\
Particles & $256^{3}$ & $128^{3}$ & $1024^{3}$ \\ 
$M_{\rm{res,DM}}\, (\rm{M_{\odot}})$ & $2082 $ &  $1.6 \times 10^{7} $ & $8 \times 10^{7} $ \\
Spatial res. (proper pc) & 7.63 & 76.3 & 1000 \\
$m_{\rm{res,\star}}\, (\rm{M_{\odot}})$ & $2.3 \times 10^{2}$ &  $7.7 \times 10^{3}$ & $2 \times 10^{6}$\\
\hline
\end {tabular}
\end{center}
\label{comp_simu}
\end{table}

\subsection{Star formation and physical processes}
In all simulations, star formation is allowed in cells with a gas density exceeding the threshold $\rho_0$, which is $30\, \rm  H\, cm^{-3}$ for Tiny, $1 \, \rm H\, cm^{-3}$ for Chunky, and $0.1 \, \rm H\,cm^{-3}$ for Horizon-noAGN, and modeled with a Kennicutt-Schmidt law:

\begin{equation}
\dot{\rho}_{\star}=\epsilon_{\star}\frac{\rho}{t_{\rm{ff}}},
\end{equation}
with $\dot{\rho}_{\star}$ the star formation rate density, $\epsilon_{\star}=0.02$ the star formation efficiency (constant with redshift), $\rho$ the density of the gas and $t_{\rm{ff}}$ the free-fall time of the gas.
Stars are created with a Poisson random process calculating the probability to form $N$ stars with a mass resolution of $m_{\rm{res,\star}}=2.3 \times 10^{2} \, \rm{M_{\odot}}$ for Tiny,  $7.7 \times 10^{3} \, \rm{M_{\odot}}$ for Chunky, and $2 \times 10^{6} \, \rm{M_{\odot}}$ for Horizon-noAGN \citep{rasera&teyssier06}.
To mimic reionization, heating from an uniform UV background is added \citep[following][]{Haardt96}, taking place after redshift $z=8.5$ for the simulations Chunky and Tiny, and $z=10$ for Horizon-noAGN.
Cooling is modeled with  the cooling curves of \citet{sutherland&dopita93}, the gas cools through H, He, and metals.  Modeling the metallicity as a passive variable makes it easily trackable over the gas flow through redshift evolution. Physical processes, such as SN explosions and star formation, modify and redistribute the metallicity over neighboring cells. An initial zero metallicity is assumed for Chunky.
However, for the smallest box size simulation, as we resolve mini-halos below the threshold for atomic hydrogen cooling, we use a metallicity floor of $10^{-3} \, \rm{Z_{\odot}}$ in order to mimic cooling by molecular hydrogen.
The same metallicity background is employed in Horizon-noAGN. The cooling by metals also mimics here cooling by molecular hydrogen, therefore allowing the formation of stars in mini-halos. 
The gas follows an adiabatic equation-of-state (EoS) for monoatomic gas with adiabatic index $\gamma=5/3$, except at high gas densities $\rho>\rho_0$, where we use a polytropic EoS to increase the gas pressure in dense gas in order to limit excessive gas fragmentation by mimicking heating of the interstellar medium from stars~\citep{springel&hernquist03}:
\begin{equation}
T=T_{0}\left(\frac{\rho}{\rho_{0}}\right)^{\kappa-1}
\end{equation}
with $T$ the gas temperature, $T_{0}$ the temperature threshold, $\rho_{0}$ the density threshold, and $\kappa$ the polytropic index of the gas.
We use $\rm \kappa = 1.6 $ for the polytropic index, and $\rm T_{0}=10^{3}\, K$ for Chunky, and $\rm T_{0}=10^{2}\, K$ for Tiny. For Horizon-noAGN, $\rm \kappa = 4/3 $ and $\rm T_{0}=10^{3} \,K$ have been used.

\subsection{Supernova feedback}
\label{sec:sec_SN}
Because metallicity and star formation are both a direct product of how SN feedback is modeled in hydrodynamical simulations, and are of the crucial importance when studying the direct collapse scenario, we test two different SN feedback implementations. We model SNe type II assuming a Chabrier initial mass function, where 20\% of the mass fraction of stars end up their life in type II SNe, and release $10^{50} \,\rm{erg\, M_{\odot}^{-1}}$, and return metals with a yield of $0.1$.
We use a weak ``thermal" SN feedback which releases only internal energy in the neighboring cells \citep{2008A&A...477...79D}. We contrast this with a more efficient SN feedback,``delayed cooling" which accounts for the energy released by the explosion that can be stored by non-thermal processes, such as unresolved turbulence, cosmic rays, magnetic fields, etc.  These processes will dissipate their energy on potentially longer timescales, defined as the dissipative time $t_{diss}$. In order to mimic the energetic and pressure enhancement by the non-thermal component, in the delayed cooling implementation gas cooling is prevented in gas cells where the non-thermal energy component (or non-thermal velocity dispersion~$\sigma_{\rm{NT}}$) is larger than some user-defined threshold~\citep{Stinson2006,2013MNRAS.429.3068T}. We adopt the implementation of \citet{2015MNRAS.452.1502D} in order to match the values of $t_{\rm diss}$ and $\sigma_{\rm NT}$ to that required for the blast wave to propagate over a Jeans length (4 high-resolution cells).
Finally, $\rm{M_{gmc}}$ is the stellar mass exploding in SN, which we take equal to $10 \times m_{\rm{res,\star}}$ for all the simulations.  
For Chunky we use the parameters: $\rm{M_{gmc}}=7.7\times 10^{4}\, \rm{M_{\odot}}$, $\rm{\sigma_{NT}}=65 \, \rm{km \, s^{-1}}$, $\rm{t_{diss}}=4.6 \, \rm{Myr}$. For Tiny we use: $\rm{M_{gmc}}=2.5 \times 10^{3} \, \rm{M_{\odot}}$, $\rm{\sigma_{NT}}=38 \, \rm{km \, s^{-1}}$, $\rm{t_{diss}}=0.59\, \rm{Myr}$. Horizon-noAGN includes a ``kinetic feedback" with a strength intermediate between these two implementations.

\subsection{Halo catalog and merger trees}
We construct catalogues of halos using the AdaptaHOP halo finder~\citep{Aubert+04}, which uses an SPH-like kernel to compute densities at the location of each particle and partitions the ensemble of particles into sub-halos based on saddle points in the density field. Halos contain at least 20 particles. We study the individual evolution of halos by building merger trees using the code {\sc TreeMaker} developed by~\citet{Tweed+09}.  

\section{Method}
We post-process the identification of all the regions, in a given simulation, which are eligible for the formation of a DCBH. In large cosmological simulations the main difficulty is to capture both large scales to have statistics and very small scales where the collapse of gas and the formation of a massive central object can be resolved. Several physical processes, playing a crucial role for the direct collapse scenario, such as  star formation,  metal-enrichment, depend on the simulation resolution. In order to take this into account, we have run our DCBH finder code on three different simulations. Therefore we are covering a large range of resolutions and volumes.

According to the direct collapse scenario, metal-free/metal-poor halos with mass $>10^{7}-10^8 \msun$ may host DCBHs under specific conditions. An inflow rate (higher than $0.1\, \rm{M_{\odot}/yr}$) of gas at the center of the halo can lead to the formation of a supermassive star-like object in the nucleus. The star can then collapse and form a $10^{5}-10^{6}\, \rm{M_{\odot}}$ BH.
In order for the Jeans mass to remain sufficiently high to form only one very massive object, efficient cooling by molecular hydrogen or metals must be prevented. Therefore metal-free conditions and a strong photo-dissociating flux are required.

A12 and A14 model the Lyman-Werner radiation as the sum of a spatial varying component of the radiation produced by young stars and of a background component. The background component is described by:

\begin{equation}
\rm{J_{LW,bg,II}}=0.3 \left( \frac{1+z}{16}\right)^{3}\left(\frac{\dot{\rho}_{\star,II}}{10^{-3}\,\rm{M_{\odot}\,yr^{-1}Mpc^{-3}}}\right) 
\end{equation}

\noindent in units of $10^{-21}\, \rm{erg \, s^{-1}\, cm^{-2}\, Hz^{-1}\, sr^{-1}}$, with  $\dot{\rho}_{\star,II}=10^{-3}\, \rm{M_{\odot}\,yr^{-1}Mpc^{-3}}$ for the star formation rate (constant with time). The background radiation intensity, however, is negligible compared to the local radiation intensity: 

\begin{equation}
\rm{J_{LW,local,II}}=3 \sum_{i,stars \leqslant 5Myr}\left( \frac{r_{i}}{1\, \rm{kpc}}\right)^{-2}  \left(\frac{m_{i}}{10^{3}\,\rm{M_{\odot}}}\right)
\end{equation}
\noindent in units of $10^{-21}\, \rm{erg \, s^{-1}\, cm^{-2}\, Hz^{-1}\, sr^{-1}}$, with $\rm{r}$ the distance between the source and the region where we compute the radiation and $\rm{m}$ the mass of the star. In the following, $\rm{J_{LW, crit}}$ refers to the critical value of $\rm{J_{LW,local,II}}$ in units of $10^{-21}\, \rm{erg \, s^{-1}\, cm^{-2}\, Hz^{-1}\, sr^{-1}}$, which we simplify in $\rm{J_{LW}}$.  In this study, as we will consider radiation intensity thresholds much above the background level, we do not include the background component to the radiation intensity, which would be negligible compared to the spatially varying component. We have also not included radiation produced by Pop~III stars, as their contribution to DCBH formation is highly subdominant compared to the Pop~II population (see A12 and A14). 

In this work, we do not include self-shielding by molecular hydrogen, self-shielding of halos could decrease the number density of DCBH regions we find in the following sections. 
We mimic reionization by adding a uniform UV background, and we do not add any additional ionizing radiation in the post-processing analysis. \citet{Johnson14}, using a zoom-in simulation of a $10^{7} \,\rm{M_{\odot}}$ halo, show that ionizing radiation has an effect only before the beginning of the collapse of the inner part of halos, and gas infall and accretion in the center can be decreased, which delays collapse, and favours molecular hydrogen formation. In this case, star formation instead of DCBH formation should occur. \citet{Johnson14}, however, show that the fraction of halos affected by this process is very small.

The critical value required for DCBH depends of the spectrum of the stellar population, \citet{Wolcott2011} argue a $\rm{J_{LW, crit}}=1000$ for a Pop III population, \citet{Shang2010} show that the critical value ranges in $J\rm{_{LW, crit}}=30-300$ for a Pop II population. A strong radiation is needed in order to not only dissociate the molecular hydrogen in the outer parts of the halo, but also in the center. The critical values cited just above are thought to be sufficient to bring the molecular fraction down to $10^{-8}$ in 1D simulations. The value of the critical radiation $\rm{J_{LW, crit}}$ is more likely to be spread on a distribution of possible values rather than equal to a fixed single value (derived from a fixed temperature black body), as shown by \citet{Sugimura14,Agarwal2014,Agarwal2015}. However in order to be able to compare our results on the number density of DCBH regions to previous literature, we consider a critical value of the radiation intensity of either $\rm{J_{LW, crit}}=30$ \citep{Agarwal2012SAM, Agarwal2014}, $\rm{J_{LW, crit}}=100$ or $\rm{J_{LW, crit}}=300$. These values are lower compared to those required by high-resolution 3-D cosmological simulations \citep{Regan2014,Latif2014ApJ,LatifXray2015}, which require $\rm{J_{LW, crit}}>500-1000$. Inclusion of X-rays also increases the critical flux \citep{2015MNRAS.450.4350I}, but the net effect is still unclear \citep{LatifXray2015}. 
It will be clear in the following that if we were to consider such values we would not form any DCBH in our volumes. In summary, what we obtain is an upper limit to the number of DCBHs, under optimistic conditions.

\section{Impact of SN feedback on metallicity and star formation}

\begin{figure}
\centering
   \includegraphics[width=6.8cm]{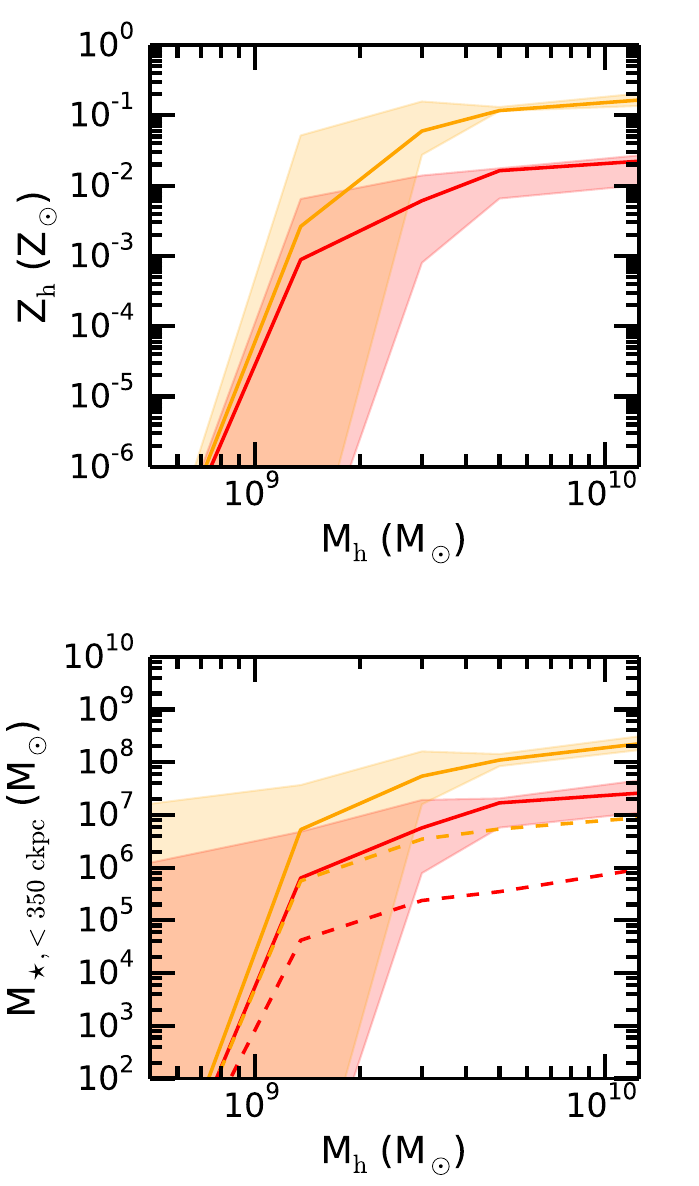}
\caption{Top panel: Median of the average metallicity of halos in the two Chunky simulations, at redshift $z=7.33$, binned in mass. Bottom panel: median stellar mass enclosed in a sphere with a radius of 350 ckpc around halos (solid lines, red for the simulation with the delayed cooling SN feedback and yellow for the thermal SN feedback). Dashed lines include only stellar mass in young stars (age $\leqslant  5$~Myr). Shaded areas extend to the $1\sigma$ values. The spread in the metallicity, as well as in the stellar mass, is due to very isolated halos. thermal feedback decreases the probability of having DCBH regions because the number of halos polluted by metals is higher, but it also favours it because the amount of radiation produced by young stars is higher.}
\label{fig:met_stars}
\end{figure}

Metals are created by stars and spread by SN explosions. Therefore SN feedback implementation has a strong impact on the metal-enrichment of the intergalactic medium, and on the number density of BH seeds formed. SN feedback also regulates star formation, therefore modulates the formation of the young stars that can provide Lyman-Werner radiation. 

The simulation Tiny allows us to resolve the detailed expansion of the metal-enriched bubbles, but the size of the simulation (1 cMpc) is  too small to see, statistically, the impact of SN metal enrichment on the number density of DCBHs. We use Chunky, a simulation with a side length 10 times larger (10 cMpc), to study metal enrichment and star formation rate for a significant volume of the Universe. We have run two identical simulations where only the prescription for the SN feedback is different, as described in Section 2.6. The thermal SN feedback is weaker, compared to delayed cooling. In fact, the total mass in stars in the box is about one order of magnitude larger for the thermal feedback case, at all redshifts. 

The direct collapse scenario depends on two main quantities: the metallicity of the halo, and the radiation intensity it is exposed to, which in turn depends on the mass in nearby young stars ($<5$ Myr). We compare these two quantities for the two SN feedbacks in Fig.~\ref{fig:met_stars}. We calculate a halo mean metallicity averaging over all the gas leaf cells enclosed in its virial radius, and show the median of these mean metallicity values for all the halos in the two Chunky simulations, with solid lines in Fig.~\ref{fig:met_stars} (top panel). The shaded areas represent the $1\sigma$ values of halos metallicity. Thermal feedback leads to a higher metal enrichment. This is simply due to the larger stellar mass formed when adopting the SN thermal feedback, as evident from the bottom panel of Fig.~\ref{fig:met_stars}, which represents the median stellar mass in the neighborhood of halos (solid lines), in a sphere with a 350 ckpc radius. This median stellar mass is larger with the thermal feedback.  The stellar mass in young stars ($\leqslant 5$ Myr), which contribute at a given redshift to the LW radiation, is shown as dashed lines and follows the same trend.

It is difficult to predict the global effect of SN feedback on the number density of DCBHs in a cosmological box because metal enrichment and the amount of  stellar mass in young stars have opposite effects.  Regarding the former, delayed cooling SN feedback is more favourable to the formation of DCBHs - more halos are metal-poor and therefore eligible for DCBH formation-, regarding  the latter, it is the thermal SN feedback implementation that has the advantage - halos are illuminated by stronger radiation because there are more young stars at any given time.

\section{Number density of direct collapse regions in Chunky}

\begin{figure}
\centering
   \includegraphics[width=9cm]{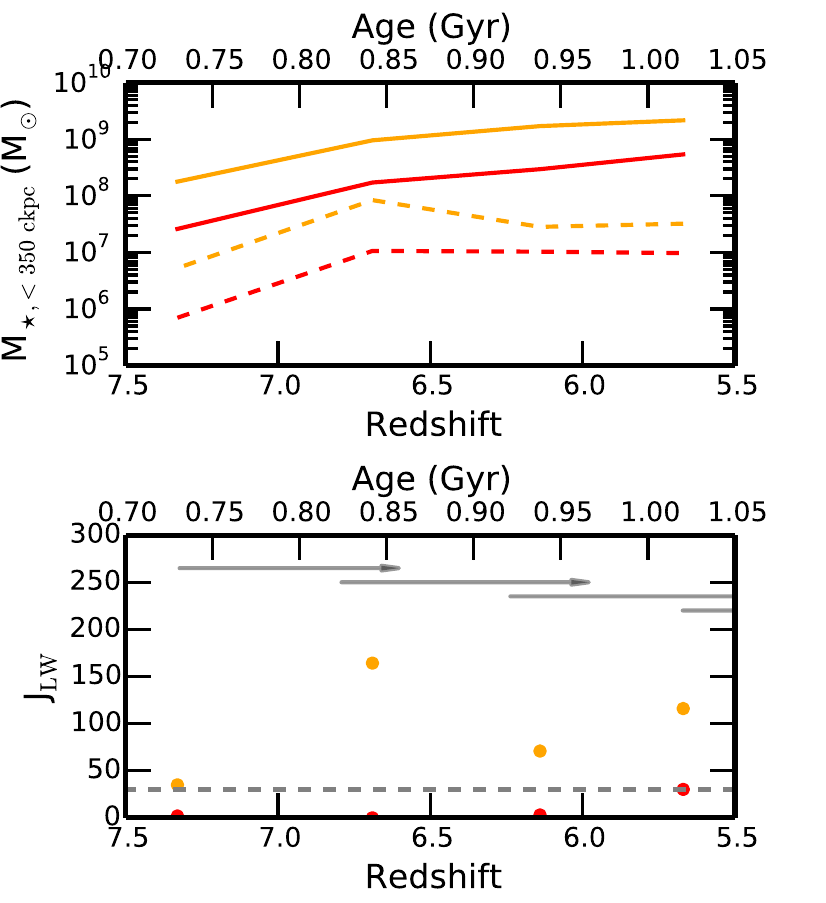}
\caption{ We compare one DCBH candidate halo in the simulation Chunky with the thermal supernova feedback (in orange) and the corresponding halo in the delayed cooling simulation (in red). Top panel:  stellar mass (solid lines),  stellar mass in young stars (dashed lines) in the neighborhood of the candidate (in a sphere of 350 ckpc radius). Bottom panel: the radiation intensity (dots) seen by the candidate halo for the two supernova feedbacks. The critical radiation intensity used in A12, $\rm{J_{LW,crit}}=30$, is shown as a dashed grey line. Arrows represent the free-fall time used in D08. If a candidate is above the critical LW radiation for the full length of an arrow, it is retained. In this case the orange dot (thermal feedback) remains a candidate, the red dot (delayed cooling feedback) does not.}
\label{fig:comp_evolution}
\end{figure}

\begin{table}
\caption{Number of direct collapse regions in the simulation Chunky with thermal SN feedback, assuming that collapse requires either 10 Myr, or the full free-fall time of the halo. For reference in Chunky with delayed cooling SN feedback, the only case that gives one candidate, at $z=7.33$,  has $\rm{J_{LW,crit}}=30$ and a collapse time of 10 Myr.}
\begin{center}
\begin{tabular}{lcccc}
\hline
\hline
Criteria & $z=10.1$ & $9.00$ & $8.09$ & $7.33$\\
\hline
$\rm{J_{LW,crit}}=30$,  10 Myr & 1 & 3 & 6 & 17\\
$\rm{J_{LW,crit}}=30$,  $\,t_{\rm{Myr, ff}}$ & 0 & 0 & 0 & 3 \\
$\rm{J_{LW,crit}}=100$,  10 Myr & 0 & 0 & 1 & 3\\
$\rm{J_{LW,crit}}=100$,  $\,t_{\rm{Myr, ff}}$ & 0 & 0 & 0 & 0\\
$\rm{J_{LW,crit}}=300$,  10 Myr & 0 & 0 & 0 & 0 \\
$\rm{J_{LW,crit}}=300$,  $\,t_{\rm{Myr, ff}}$ & 0 & 0 & 0 & 0 \\
\hline
\end {tabular}
\end{center}
\label{comp_ndcbh}
\end{table}

We now turn to identifying the number of regions which are eligible for the direct collapse scenario. These regions must fulfill three criteria: they must be metal-poor, not forming stars, and be illuminated by a LW radiation intensity higher than the threshold $\rm{J_{LW, crit}}=30$ (we take here the same value as in A12) during the whole time it takes for the collapse and DCBH formation. D08 use a free-fall time defined by:
\begin{equation}
t_{\rm{Myr, ff}}\sim 83 \left( \frac{1+z}{11}\right)^{-3/2},
\label{eq:t_ff}
\end{equation}
which assumed $\rho \sim 200 \times \bar{\rho}$.
This is the free-fall time at the virial radius. \citet{Visbal14} argue instead that the relevant collapse time is that of the gas, on scales 10\% of the virial radius, so that the collapse time should be $0.1\times t_{\rm{Myr, ff}}$. The length of the collapse in 3D high-resolution simulations of \citet{Latif2015}, indeed lasts $\sim$10~Myr. 
In this study, we consider both the redshift-dependent collapse time, defined in Eq.~\ref{eq:t_ff} (D14), and a collapse time of 10 Myr, which appears more realistic, according to recent simulation studies.

We give an example of our technique in the following. We show in Fig.~\ref{fig:comp_evolution} one concrete example of a DCBH candidate halo, with a mass of $M_{\rm{h}}=5.2 \times 10^{8}\, \rm M_{\rm{\odot}}$ at $z=7.33$: the top panel shows the evolution of the stellar mass in the environment of the halo, the total stellar mass is represented with solid lines, the stellar mass in young stars is shown with dashed lines. The corresponding radiation intensity is shown in the bottom panel with dots, the grey dashed line indicates the radiation intensity threshold of $\rm{J_{LW,crit}}=30$. The simulation with thermal feedback is shown in orange, and the simulation with the delayed cooling feedback in red. Grey arrows give an idea of the free-fall time at that redshift, computed as in D08. 
We see that when the stellar mass in young stars increases, the radiation intensity also increases, and when it decreases at redshift $z=6.6$, the radiation intensity also decreases. 

Using this technique, only considering the radiation criterion, for the thermal SN feedback case, 17 regions are identified as potential DCBH sites, at $z=7.33$. The maximum value of the radiation intensity at that redshift is $\rm{J_{LW}}=162.72$. Then we look back in time using merger trees made with TreeMaker \citep{Tweed+09}, in order to check if these candidates have been illuminated by a sufficiently high radiation intensity for at least one collapse time, and how the intensity varies over time. We find that only 4 of the candidates are illuminated by a radiation intensity above the critical value for at least one collapse time, and therefore are still flagged as DCBH candidates. 
Finally we check the metallicity of the candidates, and find that only one region is not polluted by metals, and 2 regions are partially polluted, which leads to a number density $\rm{n_{DCBH}} \leqslant 3 \times 10^{-3} \rm{cMpc^{-3}}$. Zero candidates are found for $\rm{J_{LW,crit}}=100$ and $\rm{J_{LW,crit}}=300$.

In the previous paragraph, we adopted as collapse time the free-fall time of the halo, defined by Eq.~\ref{eq:t_ff}. However, if we now use a collapse time of 10 Myr, as suggested by \citet{Visbal14},  for the same $\rm{J_{LW, crit}}=30$, we find: 1 halo at $z=10.1$ with  $Z <10^{-3.5} \,\rm{Z_{\odot}}$, 3 regions at $z=9.00$ (all with $Z=0$), 6 regions at $z=8.09$ (3 regions with $Z=0$, and 3 regions with $Z <10^{-3.5} \,\rm{Z_{\odot}}$), 17 regions at $z=7.33$  (7 regions with $Z=0$, and 10 regions with $Z <10^{-3.5}\, \rm{Z_{\odot}}$). All the regions are in halos with $M_{\rm{h}}>10^{8} \, \rm{M_{\odot}}$.  With a radiation intensity threshold of $\rm{J_{LW,crit}}=100$, the numbers decrease  to 1 at $8.09$, and 3 at $z=7.33$, no DCBH halo is found at $z=10.1$ or $z=9.00$. Zero candidates are found for $\rm{J_{LW,crit}}=300$.

We repeat the same exercise for the simulation with delayed cooling SN feedback, and find only one candidate for the DCBH process, for the case with $\rm{J_{LW, crit}}=30$, and assuming a collapse time of 10~Myr. No candidates are found in the other cases. The maximum radiation intensity is $\rm{J_{LW}}=30.33$. In conclusion, the  lower stellar mass in young stars in the delayed cooling simulation directly impacts the direct collapse scenario, by strongly decreasing the number of regions illuminated by a sufficient radiation intensity, and therefore decreasing the number density of DCBHs. The number density of DC haloes found at a given redshift (not the cumulative number density) in the two Chunky boxes is shown in Fig.~\ref{fig:number_density}, with orange symbols when using for the collapse time the free-fall time of the halo, violet symbols for a collapse time of 10 Myr, and the number of candidates is reported in Table~\ref{comp_ndcbh}. 

As a note, no DCBH is found in Tiny, for any subset of criteria. Tiny is the only simulation where we resolve minihalos, and \citet{2001ApJ...548..509M,2008ApJ...673...14O} show that a small amount of LW flux is able to delay or temporarily shut down the formation of PopIII stars. As we consider all stars as PopII stars in this work, we are overestimating the LW radiation, thus obtaining optimistic results for the occurrence of DCBHs. If we included the suppression of PopIII star formation through LW flux, this would decrease the probability of DCBH formation. Since, under optimistic assumptions, we find zero DCBH candidates in Tiny, including the suppression of Pop III star formation in pristine minihaloes would not change this null result. 

Small simulation boxes only allow us to derive a number density of DCBH regions for $\rm{J_{LW,crit}}=30$. This number density appears to be sightly lower than the values derived by D14 or A12, and even more so compared to A14.
The differences between implementations causing these discrepancies are analyzed and discussed in Section 7. Reionization and the minimum halo mass considered also play a role. All the regions that we describe above are in halos with $\rm{M_{h}} > 10^{8} \, \rm{M_{\odot}}$. In Chunky we do not resolve smaller halos that could be still metal-poor, and available to form a DCBH region, neither to capture the early metal-enrichment of the DCBH regions we have identified. 
A12 discuss a case with a reionization feedback from hydrogen-ionizing photons, where they suppress star and BH formation in halos with $\rm{V_{c}} < 20 \,\rm{km \,s^{-1}}$. The halo mass corresponding to this $\rm{V_{c}}$ is comparable to the minimum halo mass we resolve, therefore this should be the case with the best match. However they use an escape fraction of 0.5, which differs from the other studies and makes a direct comparison problematic. A12 provide a case ``esc1.0'' and a case ``esc0.5'' without the reionization feedback, from this two cases we infer that an escape fraction of unity lead to a number density higher by a factor $\sim$ 3 compared to the case of an escape fraction 0.5 (this analysis adopted amended data from A12, provided by Bhashar Agarwal, private communication). To obtain an estimate of what A12 would obtain for a case including reionization, but with an escape fraction of 1.0, we re-scale the case presented in A12, for ``esc0.5'' and accounting for the reionization feedback, to a case of escape fraction of unity. The corresponding number densities, from $2.7\times 10^{-3}$ to $9.0\times 10^{-3} \, \rm{cMpc^{-3}}$, are in good agreement with the number densities derived from our Chunky simulations.\\

\section{Horizon-noAGN simulation: Can direct collapse explain the BHs powering $z=6$ quasars?}

\begin{figure*}
\centering
   \includegraphics[width=\textwidth]{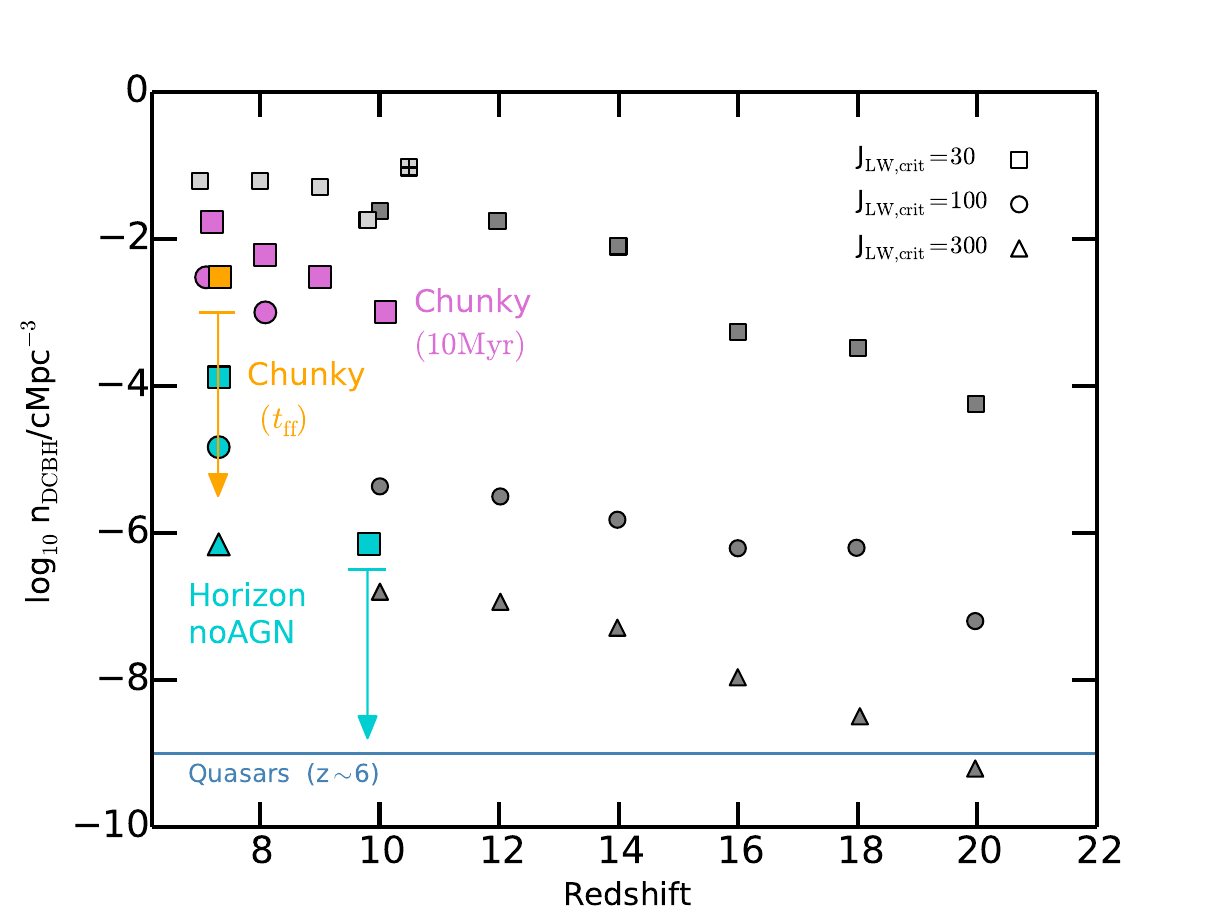}
\caption{Number density of halos that can host a DCBH, at a given redshift. To be consistent with previous literature, we show the number density of DCBH regions at a given redshift, this is not a cumulative number density in the sense that we do not add the regions found at higher redshifts. In grey we show the number density from previous studies. Symbols represent different radiation intensity thresholds. Squares:  $\rm{J_{LW,crit}}=30$, circles:  $\rm{J_{LW,crit}}=100$, triangles:  $\rm{J_{LW,crit}}=300$. The light grey crossed square at $z=10.5$ is from A14 (post-processing of a hydrodynamical simulation), the light grey squares in the range $z=10-7$ are from A12 (semi-analytical study, values from B. Agarwal, private communication), dark grey squares are the results of D14 (analytical). We do not show the number density derived in D08, which is in the range $10^{-6}-10^{-8} \, \rm{cMpc^{-3}}$, similar to the one derived in D14. The orange square shows the number density for Chunky (10 cMpc side box), for the thermal SN feedback (0 regions are identified in the simulation employing the delayed cooling SN feedback), for halos that are illuminated by a high radiation intensity for at least their free-fall time. The purple squares and circles show the number density for Chunky, for halos that are illuminated by a high radiation intensity for at least 10 Myr. The blue squares, circle and triangle represent the large-scale cosmological simulation Horizon-noAGN (142 cMpc side box).}
 \label{fig:number_density}
\end{figure*}

We now turn to exploring the number density of BHs in the Horizon-noAGN simulation. This simulation allows us to look at  DCBH formation without being biased by the feedback of pre-existing BHs, compared to the sister simulation, Horizon-AGN, which includes BH accretion and feedback. The main advantage of the simulation is that the large box (142 cMpc side length) allows some statistics, at the expense of spatial and mass resolution.  We use the same method as for the set of Chunky simulations to compute the radiation intensity illuminating halos. 

We first flag all halos illuminated by a radiation intensity higher than $\rm{J_{LW,crit}}=30$. We identify 8 regions at redshift $z=9.83$, and 814 regions at $z=7.31$. Since we are trying to estimate whether DCBHs can explain the number density of $z>6$ quasars, we do not need to proceed to lower redshift (the next available output is at $z=5.87$). This simulation includes a metallicity floor, therefore we rescaled the metallicity accordingly. Halos with a mean metallicity lower than $10^{-3.5}\rm{Z_{\odot}}$ are kept as DCBH candidates. We are left with 2 regions at redshift $z=9.83$, 373 regions at $z=7.31$. 
For a radiation intensity of  $\rm{J_{LW,crit}}=100$  we identify 2 regions at redshift $z=9.83$ and 153 at $z=7.31$.  Including the metallicity criterion ($Z  \leqslant 10^{-3.5} Z_{\rm{\odot}}$), there are 0 regions at redshift $z=9.83$, and 42  at $z=7.31$. 
For a radiation threshold of $\rm{J_ {LW,crit}}=300$ we identify 1 region at redshift $z=9.83$ and 13 regions at $z=7.31$.  Including the metallicity criterion, we find 0 regions at redshift $z=9.83$, 2 regions at $z=7.31$ ($Z  \leqslant 10^{-3.5} Z_{\rm{\odot}}$), and no region at $z=9.83$. See Table~\ref{comp_ndcbh2} for a summary. Metallicity appears to be a key parameter in the direct collapse scenario, reducing significantly the number of DCBH candidates (up to 25 \% at $z=9.83$, 45 \% at $z=7.31$). 

\begin{table}
\caption{Number of direct collapse regions in the simulation Horizon-noAGN.}
\begin{center}
\begin{tabular}{lcc}
\hline
\hline
Criteria & $z=9.83$ & $z=7.31$ \\
\hline
$\rm{J_{LW,crit}}=30$ & 8 & 814\\
$\rm{J_{LW,crit}}=30$,  $Z \leqslant10^{-3.5}  Z_{\rm{\odot}}$ & 2 & 373 \\
$\rm{J_{LW,crit}}=100$ & 2 & 153 \\
 $\rm{J_{LW,crit}}=100$,  $Z \leqslant10^{-3.5} Z_{\rm{\odot}}$ & 0 & 42 \\
$\rm{J_{LW,crit}}=300$ & 1 & 13 \\
$\rm{J_{LW,crit}}=300$,  $Z \leqslant10^{-3.5}  Z_{\rm{\odot}}$ & 0 & 2 \\
\hline
\end {tabular}
\end{center}
\label{comp_ndcbh2}
\end{table}

It is important to notice here that the time-scale between two outputs in the simulation are larger than the collapse time, either the free-fall time or 10 Myr. Therefore we are missing the time criterion of the direct collapse here: some of the candidates identified could be polluted by metals before the full collapse time has elapsed and therefore before unable to form the DCBH. The radiation intensity could also vary and not be enough to sustain molecular hydrogen dissociation for the full collapse time. Another important caveat is the resolution of the simulation, $\rm{\Delta x=1 kpc}$ and $\rm{M_{DM,res}=8 \times 10^{7} \msun}$, which does not allow us to capture the first metal-enrichment of halos, simply because we do not resolve small halos. Rather than the quantitative numbers, Horizon-noAGN allows us to explore trends in a statistical sense. 

Fig.~\ref{fig:number_density} shows the number density obtained for the simulation Chunky (in orange, and violet) and Horizon-noAGN (in light blue).  
All the grey symbols represent previous studies: the light grey crossed square is for A14, the light squares for A12 (case ``esc1.0''), dark grey squares for D14, both for $\rm{J_{LW,crit}}=30$, dark grey circles are for D14 and $\rm{J_{LW,crit}}=300$, dark grey triangles for D14 and $\rm{J_{LW,crit}}=300$. We do not show the number density derived in D08 ($10^{-6}- 10^{-8} \, \rm{cMpc^{-3}}$), because similar to the one obtained in D14.\\

In Horizon-noAGN, considering the metallicity threshold $Z \leqslant10^{-3.5} \, Z_{\rm{\odot}}$, we find $n_{\rm{DCBH}} \simeq 10^{-7} \, \rm{cMpc^{-3}}$ at $z=9.83$, and $n_{\rm{DCBH}} \simeq 10^{-4}\, \rm{cMpc^{-3}}$ at $z=7.31$ (blue squares in Fig.~\ref{fig:number_density}), for $\rm{J_{LW,crit}}=30$. At $z=7.31$, for $\rm{J_{LW,crit}}=100$,  we find $n_{\rm{DCBH}} \simeq 10^{-5}\, \rm{cMpc^{-3}}$ (blue circle in Fig.~\ref{fig:number_density}), and $n_{\rm{DCBH}} \simeq 10^{-6}\, \rm{cMpc^{-3}}$ for $\rm{J_{LW,crit}}=300$ (blue triangle in Fig.~\ref{fig:number_density}).

For Horizon-noAGN, our attempt was to see if the conditions on metallicity and on radiation intensity could be meet in more than few halos. Of course, with this simulation, we do not resolve small halos, nor the early metal-enrichment, but it gives us a first picture of the feasibility of the direct collapse scenario on a cosmological scale. For $\rm{J_{LW,crit}}=30$, we again find a number density smaller than D14, and A14. However, for the highest thresholds, $\rm{J_{LW,crit}}=100$, $\rm{J_{LW,crit}}=300$, the number density of DCBH regions are very similar to those found by D14.

Finally, we investigate the probability for massive halos at $z\sim6$ to be seeded by a DCBH. A massive halo at $z\sim6$, will host a DCBH if at least one of its progenitors was a DCBH region (namely metal-poor halos, illuminated by a strong photo-dissociating radiation intensity).  
We start by selecting all the most massive halos in the simulation Horizon-noAGN at $z=5.8$, namely the 552 halos with $\rm{M_{h}}\geqslant10^{11}\, \rm{M_{\odot}}$. We build the merger history of all these halos with TreeMaker \citep{Tweed+09}, with the two previous snapshots of the simulation, at $z=7.3$ and $z=9.8$. We compute the mean metallicity of all the progenitors of massive halos at $z=5.8$, and the photo-dissociating radiation they are illuminated by.
Of the 552 halos with $\rm{M_{h}}>=10^{11}\, \rm{M_{\odot}}$ at $z=5.8$, 155 have at least one DCBH progenitor at $z=7.3$, illuminated by a radiation with an intensity higher than $\rm{J_{LW,crit}=30}$ and with a metallicity $\rm{Z}<10^{-3.5} \, \rm{Z_{\odot}}$. We do not identify any DCBH progenitor at $z=9.8$. Therefore the fraction of massive halos which can host a DCBH is 155/552=0.28, so 28\% of the massive halos. When considering a radiation intensity threshold of  $\rm{J_{LW,crit}=100}$, only 6.5\% of the massive halos can host a DCBH, and it drops to 0.36\% for $\rm{J_{LW,crit}=300}$. See Table~\ref{comp_ndcbh3} for a summary.

In summary, about a third of the most massive halos at $z=6$ have at least one progenitor reaching both the criterion on metallicity ($\rm{Z}<10^{-3.5} \, \rm{Z_{\odot}}$) and the criterion on radiation intensity ($\rm{J_{LW,crit}=30}$). However, this fraction drops significantly down to $\sim 6\%$ for $\rm{J_{LW,crit}=100}$, and even more, down to less than 1\% for a more realistic value of $\rm{J_{LW,crit}=300}$.

\begin{table}
\caption{Percentage of massive halos with at least one DCBH progenitor, in the simulation Horizon-noAGN.}
\begin{center}
\begin{tabular}{lc}
\hline
\hline
Criteria & $\rm{Z} \leqslant10^{-3.5}$ \\
\hline
$\rm{J_{LW,crit}}=30$ & 28\% \\
$\rm{J_{LW,crit}}=100$ & 6.5\% \\
$\rm{J_{LW,crit}}=300$ & 0.36\% \\
\hline
\end {tabular}
\end{center}
\label{comp_ndcbh3}
\end{table}

\section{Comparison between hydro-dynamical simulations and (semi-)analytical models}
\begin{figure}
   \includegraphics[scale=0.4]{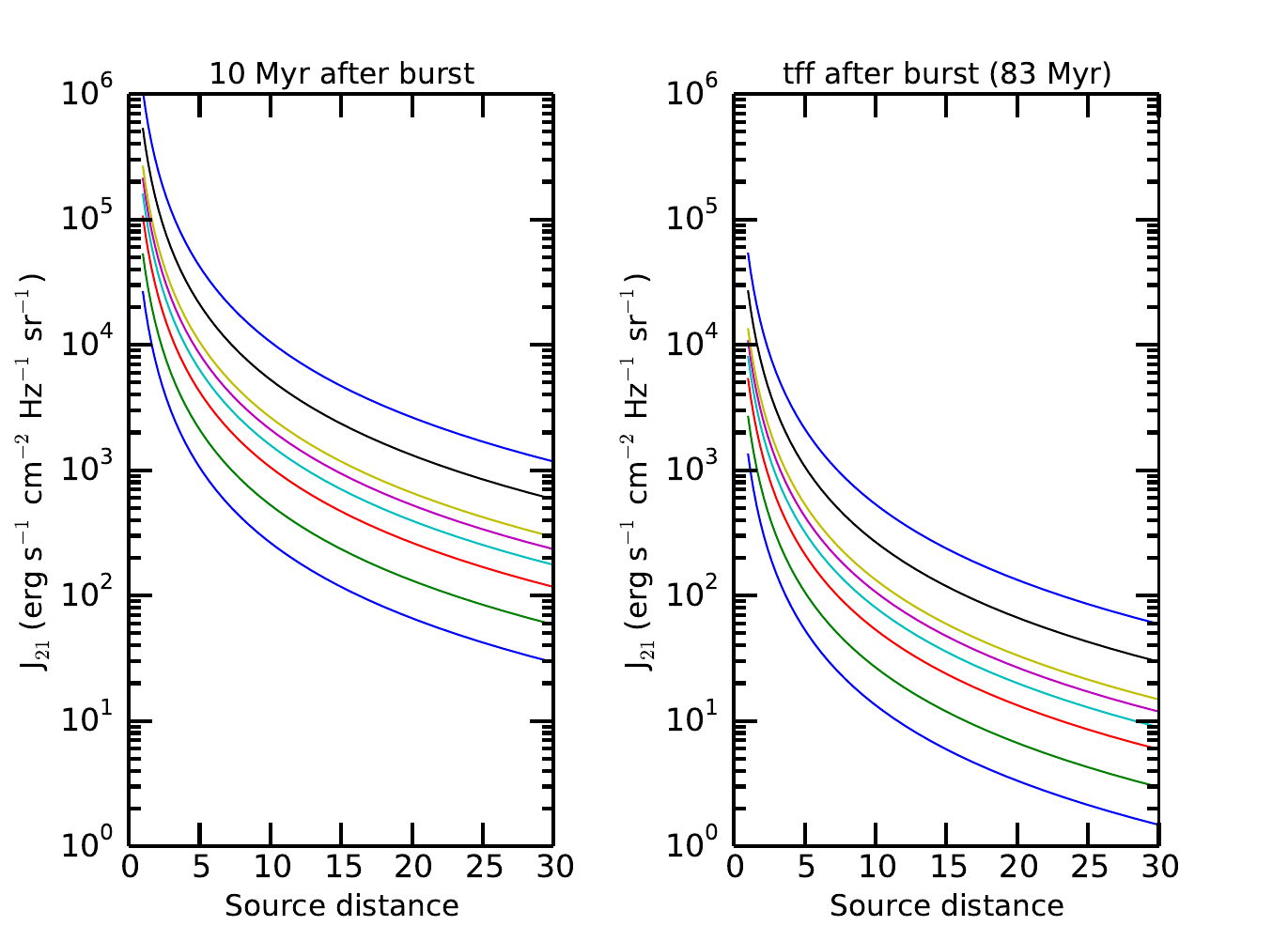}
   \caption{Radiation intensity (in $\rm{J_{LW}}$ units) provided by a source at a given distance (in proper kpc) for different sources mass at $z=10$ (halo masses of the source from top to bottom: $4 \times 10^{11}$, $2 \times 10^{11}$, $10^{11}$, $8 \times 10^{10}$, $6 \times 10^{10}$, $4 \times 10^{10}$, $2 \times 10^{10}$, $10^{10} \,\rm{M_{\odot}}$), 10 Myr after the star formation burst (left panel), and 83 Myr after the burst (right panel) using the model by  D14.}
   \label{fig:radiation_1}
\end{figure}

\begin{figure}
   \includegraphics[scale=0.42]{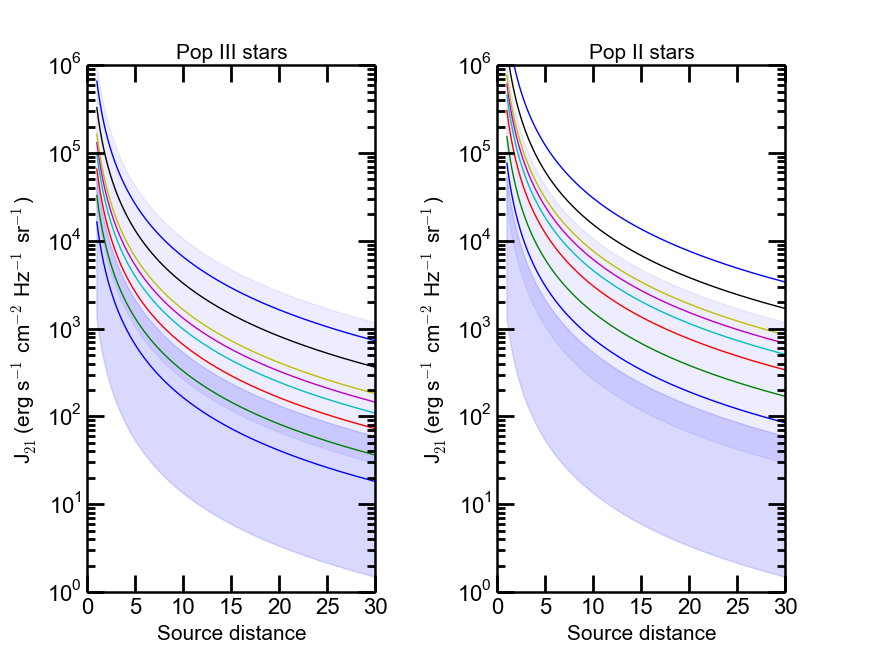}
    \caption{Radiation intensity (in $\rm{J_{LW}}$ units) provided by a source at a given distance (in pkpc) for different sources mass, at $z=10$.  The colored curves (colors as in Fig.~\ref{fig:radiation_1}) show the radiation intensity based on A14, considering a stellar population PopIII (left) or PopII (right) as the radiation source. Halo masses of the source from top to bottom: $4 \times 10^{11}$, $2 \times 10^{11}$, $10^{11}$, $8 \times 10^{10}$, $6 \times 10^{10}$, $4 \times 10^{10}$, $2 \times 10^{10}$, $10^{10}\, \rm{M_{\odot}}$. The model by D14 is shown with filled blue regions, the top region corresponds to 10 Myr after the burst, the bottom region to 83 Myr after the burst) for the same stellar masses.}
    \label{fig:radiation_2}
\end{figure}

\indent In this section, we perform a systematic comparison between different models presented in the literature.  Analytical and semi-analytical studies have the advantage of investigating with ease the impact of physical processes.  \citet{Ahn2008} show that the clustering of halos, which are the sources responsible for the photo-dissociating background, leads to local variations in the LW radiation intensity background. D14 uses a semi-analytical model to investigate the role of galactic outflows, which drive metal-enrichment in the surrounding of star-forming galaxies.  DCBH regions need to be close enough to star-forming galaxies in order to be illuminated by a high radiation intensity, whereas  the proximity with star-forming regions imply a potential metal-pollution by their galactic winds. \citet{Agarwal2012SAM} and \citet{2015arXiv150705971H} use a hybrid model where they ``paint" galaxies on dark matter only simulations, so that the clustering of halos is naturally taken into account. 

Conversely, hydrodynamical simulations have the advantage of tracking the cosmic evolution of  metal-enrichment and star formation in a more self-consistent way, where semi-analytical models need to use approximations.  However, simulations cannot resolve large and small scales at the same time:  small scale simulations allow one to follow the physical processes accurately but suffer from poor statistics (especially considering that the probability of having a DCBH is $10^{-6}$, based on the analytical estimates), while large scale simulations provide statistics but do not capture physical details/processes. For instance, \citet{LatifXray2015} use zoomed simulations of single halos to investigate different values of $\rm{J_{LW,crit}}$. A14 use a high resolution simulation with a size of 4 cMpc to estimate the density number of DCBHs. Our work covers the scale of A14, and larger scales up to 142 cMpc box size length, at degrading resolution.

We start by comparing different models for the radiation intensity coming from star-forming regions, D14 versus A14 (the latter based on A12), and then move on to the probability for halos to have a given stellar mass, the probability of being star-forming, and the probability of being metal-free.\\
\noindent In D14, the stellar mass of a dark matter halo $M_{\rm{halo}}$ is assigned as:
\begin{equation}
M_{\rm{\star}}=f_{\rm{\star}} M_{\rm{halo,gas}} = f_{\rm{\star}} \frac{\Omega_{\rm{b}}}{\Omega{\rm{m}}} M_{\rm{halo}},
\label{eq:mstar}
\end{equation}
where $f_{\rm{\star}}=0.05$ is the fraction of gas which turns into stars, $M_{\rm{halo,gas}}$ the gas mass of the halo,  $M_{\rm{halo}}$ the total mass of the halo, $\Omega_{\rm{b}}$ the baryon density and $\Omega_{\rm{m}}$ the total matter density. 
The mean production rate of LW photons per solar mass of star formation is time-dependent, where time is counted from the time $t_{\rm{Myr}}$ when a burst of star formation occurs, and expressed as:
\begin{equation}
\left<Q_{\rm{LW}} (t) \right>= Q_{\rm{0}} \left(1+\frac{t_{\rm{Myr}}}{4}\right)^{-3/2} \exp\left({-\frac{t_{\rm{Myr}}}{300}}\right)
\end{equation}
with $Q_{0}=10^{47}\,\rm{s^{-1}} \, \rm{M_{\odot}^{-1}} $.

The mean LW photon production rate is computed one free-fall time after the star formation burst (Eq.~\ref{eq:t_ff}). D14 motivate this choice by the requirement that molecular hydrogen is suppressed throughout the collapse.  The expression of $Q_{\rm{LW}}$ is a fit from STARBURST99 (which used a Salpeter IMF in the range $m_{\rm{low}},m_{\rm{up}}=1,100\,  \rm{M}_{\rm{\odot}}$, an absolute metallicity of $Z=10^{-3}$ (0.05 $\rm{Z_{\odot}}$), and a stellar mass of $10^{5} \,\rm{M}_{\rm{\odot}} $).
The mean LW luminosity density $\left< L_{\rm{LW}}(M,t) \right>$ is a function of the mean number of LW photons (given by the mean production rate of LW photons per solar masses times the stellar mass of the halo), their energy and the escape fraction of these photons. \begin{equation}
\left< L_{\rm{LW}}(M,t) \right>=\frac{h \left< \nu \right>}{\Delta \nu } \left<Q_{\rm{LW}} (t) \right> f_{\rm{esc,LW}} \left( \frac{M_{\rm{\star}}}{M_{\rm{\odot}}} \right).
\end{equation}
The efficiency of LW photons to escape from their halos is highly debated, and can depend on halo mass and stellar feedback. Using hydrodynamical zoom-in simulations of high-redshift mini-halos, \citet{Kimm14} find that the escape fraction could be close to 100\% at the epoch of reionization. \citet{Schauer15} recently showed that the escape fraction from PopIII stars can be close to 100\% in the ``far-field'' limit, but can significantly decrease by taking also into account self-shielding by atomic hydrogen. In order to be able to compare our models directly to previous works, which for the most part used an escape fraction of 100\%, we adopt the same value. \\
The flux at a distance $r$ then becomes:

\begin{equation}
\left< J_{\rm{LW}}(r,M,t_{\rm{ff}}) \right> = \frac{1}{4 \rm{\pi}} \frac{\left< L_{\rm{LW}}(M,t) \right>}{4 \rm{\pi} r^{2}} f_{\rm{mod}}(r),
\end{equation}
where the first factor $1/4\rm{\pi}$ is needed to express $\left< J_{\rm{LW}}(r,M,t_{\rm{ff}}) \right>$ in $\rm{J_{LW}}$ units ($\rm{erg \, s^{-1} \,cm^{-2}\, Hz^{-1}\, sr^{-1}}$).
 $ f_{\rm{mod}}(r)$ is used to correct the radiation intensity for the extra dimming introduced by the LW horizon \citep{Ahn2008}:

  \begin{align}
f_{\rm{mod}} (r)&=  1.7 \exp\left(-\left(\frac{r_{cMpc}}{116.29\, \alpha}\right)^{0.68}\right)-0.7 &&  \rm{if}~ r_{\rm{cMpc}}/\alpha \le 97.39\\
&=  0 &&   \rm{otherwise}.
\end{align}

\noindent In Fig. ~\ref{fig:radiation_1} we show the intensity (y axis in units of $\rm{J_{LW}}$) of the radiation coming from a source at a given distance for  different stellar masses of the source. In the left panel, we consider the radiation emitted 10 Myr after the star formation burst, while in the right one we use one free-fall time after the burst ($\sim$83 Myr at redshift z=10), as in D14. 
In Fig. ~\ref{fig:radiation_2}, the radiation intensity computed using the model of D14 is shown in blue shades for 10 Myr or 83 Myr, and is compared to the model of  A14 for the same stellar masses as in Fig.~\ref{fig:radiation_1}. Sources are considered either as Pop III stars (left panel in Fig. ~\ref{fig:radiation_2}), or PopII stars (right panel). We find that the model of  A14, considering the source either as a population of PopIII or PopII stars, roughly corresponds to the model of D14 for 10 Myr after the star formation burst, but overestimate D14 for a free-fall time (as computed with Eq.~\ref{eq:t_ff}). 

A14 model the radiation intensity in dependence only on the stellar mass of the source (considering only stars younger than 5 Myr), and the distance to the stellar source, and therefore it is not explicitly time-dependent. The time-dependence is implicit in the choice of using only stars younger that 5 Myr and in assuming a star formation history.  This differs with respect to the explicit time dependence used in D14 (10 Myr, 83 Myr) and A12. D14 and A12 provide a time-dependent model of the radiation intensity. A12 (erratum), and seemingly A14 (based on a comparison between their coefficient and the erratum of A12), used continuous star formation to model the radiation intensity. With a starburst, the emission rate of LW photons drops rapidly. With continuous star formation, it increases rapidly at first, and becomes constant after 80 Myr (cf. Fig. 2 in A12 with Fig. 1 in the erratum of A12, and see Fig. A1 in D14). The radiation intensity of A12 and A14, for continuous star formation, would be higher than the one derived by D14, who adopted a starburst, if we considered a time longer than 5 Myr (e.g., 10 Myr or 83 Myr). Over timescales longer than $\sim$ 10 Myr, however, metals generated by the stars themselves would pollute the regions irradiated by the LW flux \citep[D14,][]{2015arXiv150705971H} \citep[but see][]{2008ApJ...674..644C,2013ApJ...772..106M,2015MNRAS.452.2822S}, making the photon production ineffectual in view of the DCBH process.

We then  compare another quantity, the stellar mass per halo. D14 use a linear relation (Eq.~\ref{eq:mstar}) that overestimates the stellar mass compared to our hydrodynamical simulation. We compare the stellar mass of all halos in the Chunky simulation at $z=9$ to the theoretical stellar mass derived with D14 formalism. On average the masses of D14 are a factor $\sim 70$  larger than those in Chunky (albeit with a large scatter). On the other hand, in D14, the probability for a halo to be star-forming is set  at $\rm P_{SF}=0.1$, i.e., 10 \% of halos experience a starburst at the same given time, and contribute to the radiation intensity seen by all the other halos. This probability has a strong impact on the number density of DCBH regions, as a higher star-forming probability implies a higher radiation intensity seen by the neighboring halos. In our simulation, we define a halo as star-forming if young stars ($\leqslant 5$ Myr) are found within its virial radius.  The fraction of star-forming halos we find in Chunky is always higher than 10\%. Depending on the SN feedback model used, the fraction is between 20 and 35\% for the delayed cooling model, and between 25 and 45\% in the thermal feedback model.  

\begin{figure}
\centering
   \includegraphics[scale=0.4]{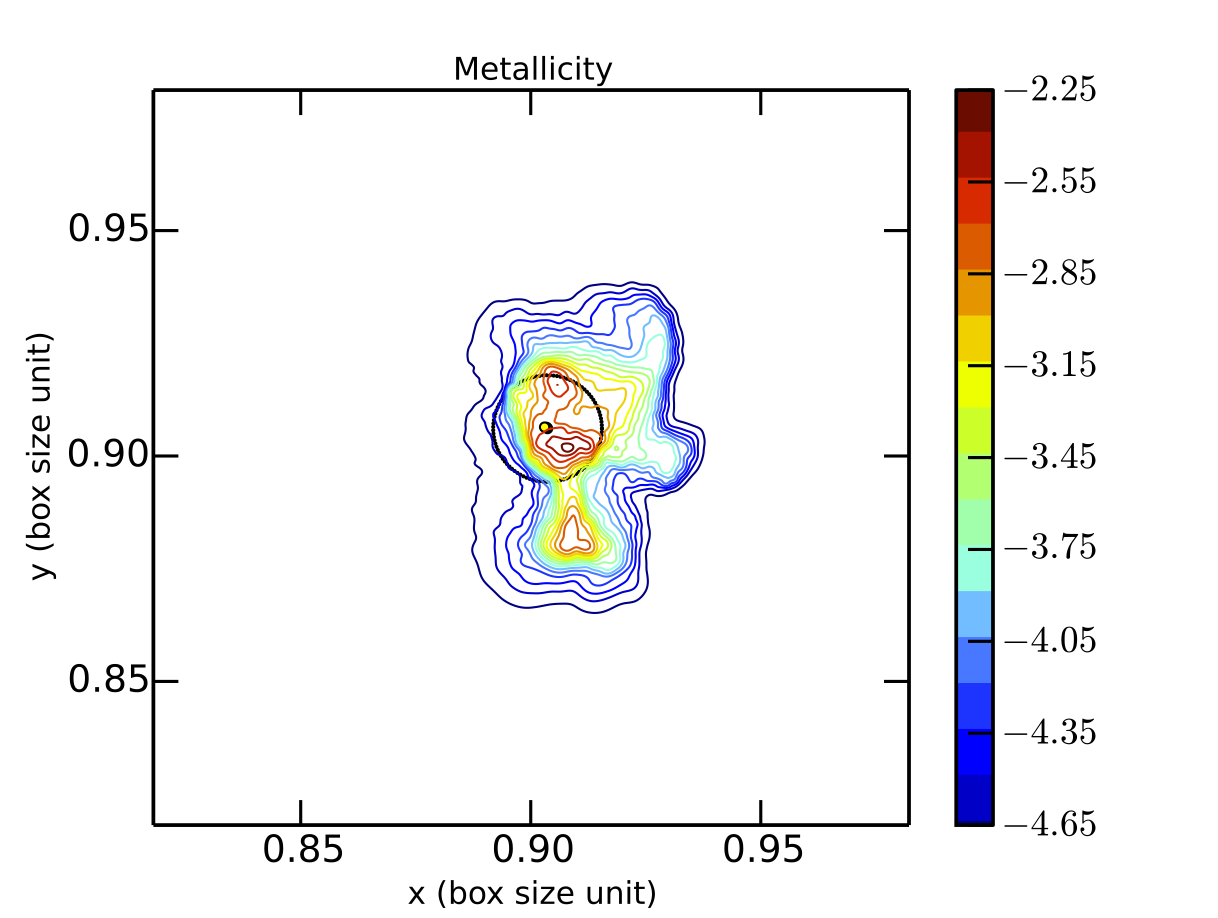}
   \includegraphics[scale=0.4]{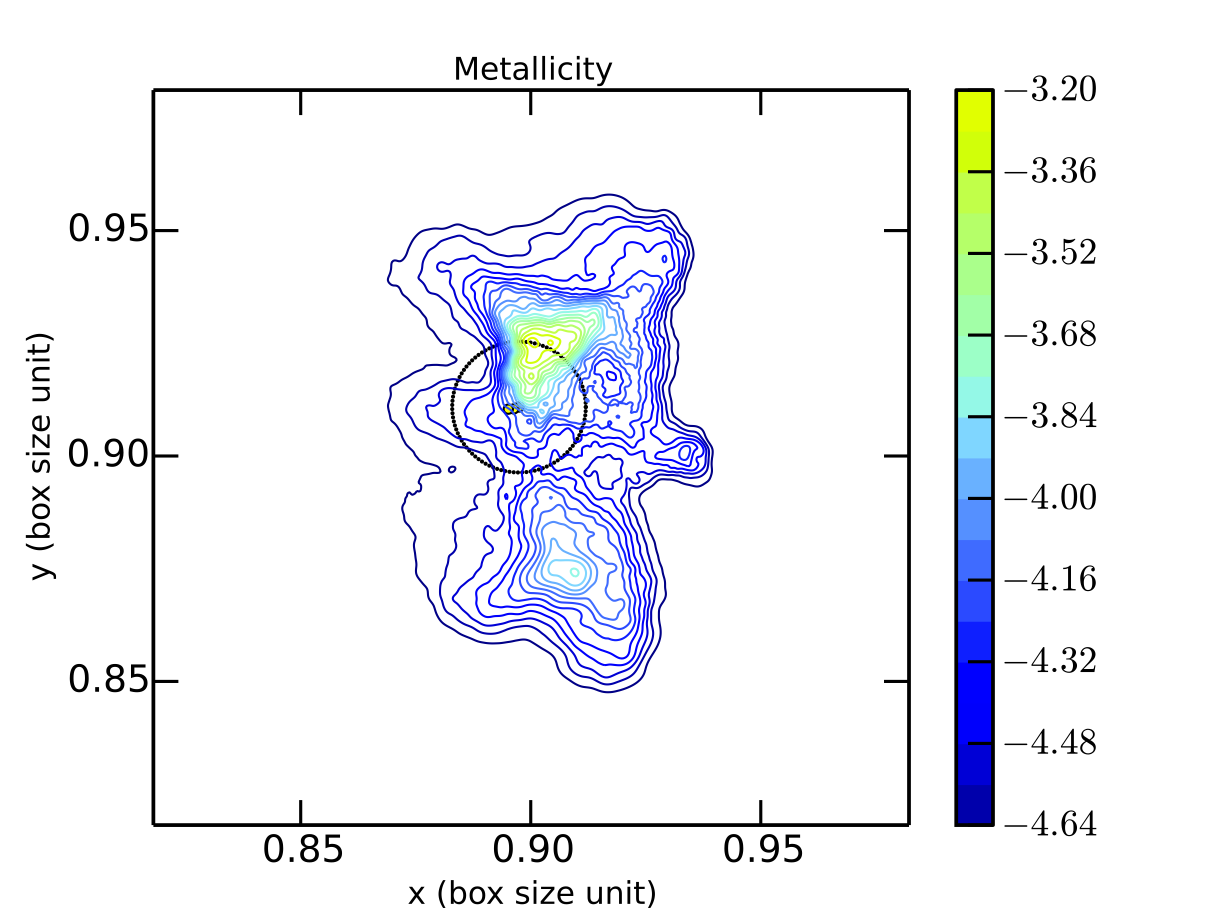}
   \includegraphics[scale=0.4]{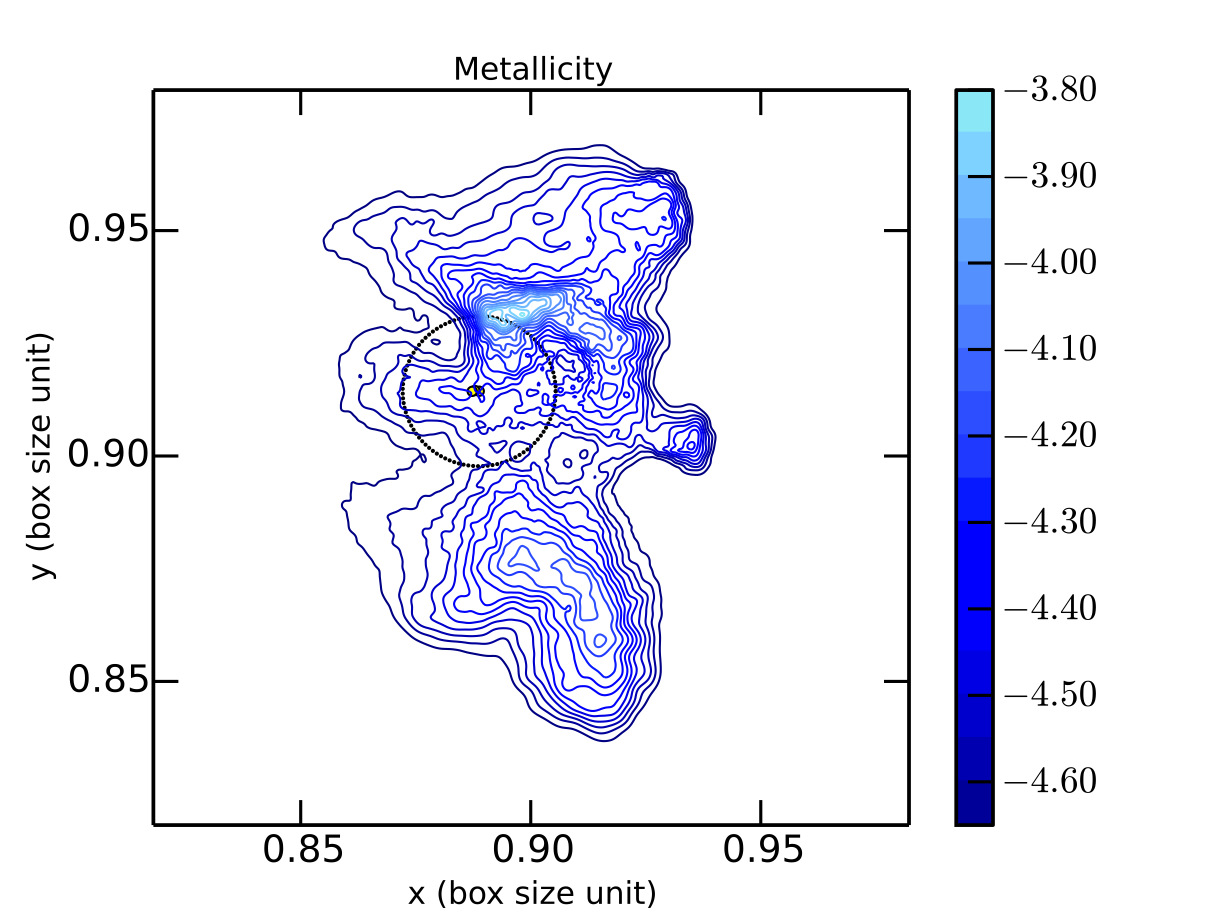}
\caption{Contour plot representing the metallicity (log10 of the absolute metallicity) around the SN explosion in the simulation Tiny at redshift $z=$10.11, 9.00,  8.09 (top to bottom). The black circle marks the analytical estimate by D14.}
\label{fig:tiny_SN}
\end{figure}

Finally, we compare the expansion of metal-polluted bubbles in the analytical framework and in simulations, using our highest resolution simulation, Tiny, which has a spatial resolution of 7.6~pc and a dark matter mass resolution of $\sim 2000 \msun$.  Fig.~\ref{fig:tiny_SN} shows the evolution of one SN bubble over three different snapshots, corresponding respectively to  $z=$10.11, 9.00, 8.09. Overlaid the metallicity contours from the simulation is the analytical estimate by D14: 
\begin{equation}
r_{\rm bubble}=3\times 10^{-2} \, \rm{kpc} \left(\frac{M_\star}{M_\odot} \right)^{1/5} n^{-1/5} \left( \frac{t}{\rm Myr} \right)^{2/5},
\label{eq:eqbubble}
\end{equation}
where $n$ is gas  60 times denser than the mean intergalactic value at redshift z, i.e. $n\sim 60 \times \Omega_b\,\rho_{\rm crit} (1+z)^3/m_p$. The expansion of the metal-polluted bubble is faster in the hydrodynamical simulation, implying that D14 underestimates the size of polluted regions compared to us. However, the bubble becomes quickly highly asymmetric, therefore the geometry of the halo-halo configuration becomes of importance. 

In summary, D14 overestimates the stellar mass, but underestimate the galaxies that can contribute radiation and the extent of metal polluted regions. The net effect is that the effects almost compensate, and explain why our results using hydrodynamical simulations are very close to the results of D14.

\section{Conclusions}
In this paper, we have addressed the formation of supermassive BHs by the direct collapse scenario. Isothermal collapse is predicted to happen in halos with minimum mass of $\sim 10^{7}-10^{8}\, \msun$ that have reached the atomic cooling threshold. To avoid the fragmentation of the gas, all the efficient coolants, namely metals and molecular hydrogen, must be absent. 
The destruction and the prevention of molecular hydrogen can be done by a strong photo-dissociating radiation coming from nearby star-forming galaxies.
A large inflow rate of gas at the center of the halo, higher than 0.1 $\msun$/yr for at least one free-fall time, is required in order to form a supermassive star-like object. 

The feasibility of the direct collapse mechanism is difficult to investigate. Zoom-in simulations investigate the collapse and the accretion properties \citep[e.g.,][]{Latif2013,Regan2014,2015MNRAS.452.1233H}, but use an artificial radiation intensity. Larger scale simulations, such as Chunky or FiBY (the simulation used for A14), instead model the spatially varying radiation intensity but are not able to follow the collapse, the accretion rate in the inner part of the halo, neither to model individually the radiation coming from each star. Employing small hydrodynamical simulation boxes has the advantage of resolving in detail different aspects of the problem (chemistry, mini-halos, early metal-enrichment, for example), but fails to present a large diversity of cosmological regions, biasing the derived number density of DC regions. 

In this paper, we use three sizes of simulation box, from a small box of 1 cMpc side length, a medium box of 10 cMpc with a set of two simulations, and a large simulation box of 142 cMpc. We model the radiation intensity coming from nearby star-forming galaxies in a similar way as D14 and A14. Our attempt was to estimate the number density of regions eligible to host a DCBH, based on the absence of efficient coolants criterion: a high enough radiation intensity to destroy molecular hydrogen and a low metallicity. The idea was not to capture all the physical processes at the same time in the same simulation, but to have a larger picture of the direct collapse scenario by using, for the first time, a suite of large scale hydrodynamical simulations. Another aim was to see if some halos were illuminated by a sufficiently high radiation intensity ($\rm{J_{LW,crit}}=100, 300$), more similar to what is obtained as critical value for collapse in 3D zoom-in simulations. 

We investigated the impact of SN feedback. We use either a weaker thermal feedback or a stronger delayed cooling feedback. Star formation is lower (by one order of magnitude) in the simulation with delayed cooling SN feedback, and more in line with the predictions of halo occupation distribution \citep{2016arXiv160509394H}. A weak SN feedback allows for more young stars, but also, consequently, for an earlier metal pollution.  Using our DCBH finder code (with $t_{\rm{ff}}$ as in D14 and $J_{\rm{LW,crit}}=30$), we do not find any DCBH regions in the simulation Chunky with the delayed cooling feedback, however we do find 3 regions for the thermal SN feedback. The absence of a strong radiation field, caused by the lower star formation rate, therefore, appears to be more important than metal pollution. Besides, in these 10 cMpc side box, we do not find any halo illuminated by a radiation intensity higher than $162.71$ in $\rm{J_{LW}}$ units, down to $z=7.33$, so no DCBH can form if $\rm{J_{LW,crit}} \gg 100$. For the delayed cooling SN feedback, the maximum value of the radiation intensity is only $\rm{J_{LW}}=30.33$.

The simulation Horizon-noAGN allows us to have a more global view of the direct collapse scenario. The simulation box is large enough to have some statistics (box side length of 142 cMpc), at the price of a lower resolution, and it includes a relatively weak SN feedback. The number density of DCBH regions in this simulation varies from $7 \times 10^{-7}$ to $10^{-4} \, \rm{cMpc^{-3}}$. We find similar results as D14, specifically for the two largest thresholds $\rm{J_{LW,crit}}=100$ and $\rm{J_{LW,crit}}=300$.
However, the number density of BHs for the threshold $\rm{J_{LW,crit}}=30$ for the Horizon-noAGN simulation is smaller than what D14 and A14 obtained for the same threshold. We do not consider a radiation background, and this can have an impact when considering low intensity thresholds such as $\rm{J_{LW,crit}}=30$. Horizon-noAGN  also allows us to investigate whether the DC scenario can explain the presence of BHs in massive galaxies at $z=6$, considered as proxies for the hosts of quasars. We find that 30\% of the halos more massive than $10^{11} \msun$ at $z=5.8$, have at least one progenitor eligible to form a DCBH for $\rm{J_{LW,crit}}=30$. This probability, however, drops abruptly below 1-10\% when considering higher thresholds ($\rm{J_{LW,crit}}=100, 300$) for the radiation intensity. 

Several approaches have been used in the last few years to determine the number density of direct collapse regions, from post-processing of hydrodynamical simulations to semi-analytical methods. These approaches derive number density which differ by several orders of magnitude (from $10^{-1}$ to $10^{-9}$ $\rm{Mpc^{-3}}$).  We perform a comparison between some of these approaches, specifically with A14 and D14, in order to understand this discrepancy. We find differences in the probability for halos to be star-forming and metal-free, for the propagation of metals in the gas, and finally in the modeling of the radiation intensity itself, which in some cases compensate to produce similar results in the number density, despite the very different single assumptions. 

In summary, we find that if DC requires $\rm{J_{LW,crit}}=300$, and a halo must be illuminated by such intense field for its full collapse, the number of DCBHs may be sufficient to explain the number of high-z quasars, based on Horizon-noAGN, but not the presence of BHs in normal galaxies. If instead either $\rm{J_{LW,crit}} \sim 30$ or a halo must be illuminated only during the collapse of the central region, then DCBHs may be common also in normal galaxies, provided that SN feedback is not very strong.

 \section*{Acknowledgments}
 We thank the referee for his/her thoughtful comments. The research leading to these results has received funding from the European Research Council under the European Community's Seventh Framework Programme (FP7/2007-2013 Grant Agreement no. 614199, project ``BLACK'').  This work was granted access to the HPC resources of TGCC under the allocations x2014046955, x2015046955 and c2015047012 made by GENCI. We thank St\'ephane Rouberoul for smoothly running the horizon cluster in IAP, where the post-processing of the simulations has been done.

\bibliography{biblio,biblio_complete}

\label{lastpage}
\end{document}